\documentclass[aps,prx,preprint,onecolumn,superscriptaddress]{revtex4-1}  
\usepackage{amsmath,amssymb,bm,bbm} 
\usepackage{graphicx}  
\usepackage{color} 
\usepackage[dvipsnames]{xcolor}
\usepackage[papersize={8.5in,11in}]{geometry}
\usepackage[colorlinks=true]{hyperref}  
\usepackage{ulem}
\usepackage[mathscr]{euscript}
\usepackage[section]{placeins}
\hypersetup{
    bookmarks=true,         % show bookmarks bar?
    unicode=false,          % non-Latin characters 
    pdftoolbar=true,        % show Acrobat
    pdfmenubar=true,        % show Acrobat 
    pdffitwindow=false,     % window fit to page when opened
    pdfstartview={FitH},    % fits the width of the page to the window
    pdfsubject={},   % subject of the document
    pdfcreator={},   % creator of the document
    pdfproducer={}, % producer of the document
    pdfkeywords={} {} {}, % list of keywords
    pdfnewwindow=true,      % links in new window
    colorlinks=true,       % false: boxed links; true: colored links
    linkcolor=magenta, %red,          % color of internal links (change box color with linkbordercolor)
    citecolor=blue,        % color of links to bibliography
    filecolor=magenta,      % color of file links
    urlcolor=blue           % color of external links
} 

\usepackage{amsfonts}
\usepackage{upgreek}
\usepackage{slashed}
\usepackage{latexsym}

\newcommand{\beq}{\begin{equation}}
\newcommand{\eeq}{\end{equation}}
\def\bea{\begin{eqnarray}}
\def\eea{\end{eqnarray}}
\newcommand{\Tr}{\text{Tr}}
\newcommand{\nn}{\nonumber \\}

\newcommand{\equref}[1]{Eq.~(\ref{#1})}

\usepackage{braket}
\usepackage{comment}
\usepackage{slashed}
\usepackage{subcaption}

\begin{document}

%\preprint{arXiv:xxxx}

\title{Phases of SU(2) gauge theory\\ with multiple adjoint Higgs fields in 2+1 dimensions}

\author{Harley D. Scammell}
\affiliation{Department of Physics, Harvard University, Cambridge, MA 02138, USA}

\author{Kartik Patekar}
\affiliation{Department of Physics, Indian Institute of Technology, Powai, Mumbai-400076, India}

\author{Mathias S. Scheurer}
\affiliation{Department of Physics, Harvard University, Cambridge, MA 02138, USA}

\author{Subir Sachdev}
\affiliation{Department of Physics, Harvard University, Cambridge, MA 02138, USA}

\date{\today
\\
\vspace{0.4in}}

\begin{abstract}
A recent work (arXiv:1811.04930) proposed a SU(2) gauge theory for optimal doping criticality in the cuprate superconductors. The theory contains $N_h$ Higgs fields transforming under the adjoint representation of SU(2), with $N_h=1$ for the electron-doped cuprates, and $N_h=4$ for the hole-doped cuprates. We investigate the strong-coupling dynamics of this gauge theory, while ignoring the coupling to fermionic excitations. We integrate out the SU(2) gauge field in a strong-coupling expansion, and obtain a lattice action for the Higgs fields alone.
We study such a lattice action, with O($N_h$) global symmetry, in an analytic large $N_h$ expansion and by Monte Carlo simulations for $N_h=4$ and find consistent results. 
We find a confining phase with O($N_h$) symmetry preserved (this describes the Fermi liquid phase in the cuprates), and Higgs phases (describing the pseudogap phase of the cuprates) with different patterns of the broken global O($N_h$) symmetry. One of the Higgs phases is topologically trivial, implying the absence of any excitations with residual gauge charges. The other Higgs phase has $\mathbb{Z}_2$ topological order, with `vison' excitations carrying a $\mathbb{Z}_2$ gauge charge. We find consistent regimes of stability for the topological Higgs phase in both our numerical and analytical analyses.
\end{abstract}

\maketitle
%\tableofcontents

\section{Introduction}
\label{sec:intro}

A previous study of a 2+1 dimensional cuprate gauge theory, developed in Ref.~\onlinecite{SSST19}, fractionalised the spin density wave (SDW) order parameter by going to a rotated reference frame in spin-space and obtained a theory with of Higgs fields with multiple ($N_h$) flavors which are charged under an emergent local $SU(2)$ gauge field. 
The Higgs fields also transform under the lattice space group and time reversal; consequently these symmetries can be broken in the Higgs phase. It was found that the symmetry breaking transitions associated with these Higgs fields lead to a variety of order parameters -- constructed as gauge-invariant bilinear or trilinear combinations -- which are consistent with the symmetry breaking patterns observed in experiments on cuprates near optimal doping. 
Further, upon considering electronic degrees of freedom coupled to the Higgs fields, a rather natural description of the pseudogap phase emerged \cite{SSST19}.

In this paper, we wish to consider the strong-coupling dynamics of the SU(2) gauge theory in more detail. Apart from the Higgs phase where the Higgs fields are condensed, there can also be a confining phase where there are no excitations associated with the Higgs fields, and 
the electronic degrees of freedom resume normal Fermi liquid behaviour. Hence, in this description, the pseudogap is associated with the Higgs phase, and Fermi liquid  with the confined phase. 

Moreover, the pseudogap/Higgs phase can have a {\it topological} structure beyond that associated with broken global symmetries. This structure is associated with any gauge group left unbroken by the Higgs condensate \cite{FradkinShenker}, and is also tied to the pattern of broken global symmetry. It was found that, depending upon parameters, the Higgs condensate could break the SU(2) gauge symmetry down to U(1) or $\mathbb{Z}_2$. The U(1) gauge field confines in 2+1 dimensions, and so the U(1) case is ultimately topologically trivial. However, the $\mathbb{Z}_2$ case leads to $\mathbb{Z}_2$ topological order \cite{NRSS91,Wen91}, with deconfined excitations carrying $\mathbb{Z}_2$ electric and magnetic gauge charges. Specifically, the $\mathbb{Z}_2$ magnetic charges are carried by vortex configurations (`visons') in the Higgs fields, while the $\mathbb{Z}_2$ electric charges are carried by gapped spinons excitations.

We note that an earlier study \cite{Kibble02} of a 2+1 dimensional SU(2) lattice gauge theory with a single ($N_h=1$) adjoint Higgs field also considered the case where the Higgs phase breaks the SU(2) down to U(1) \cite{GG72}. In this case, the confining and Higgs phases were found to be continuously connected, and the theory has only one phase and no phase transition. However, in our case the topologically trivial Higgs phase does break global symmetries for $N_h>1$, and so even the trivial Higgs and confining phases remain separated by a phase transition. 

The objective of the present work is to study the strong-coupling dynamics of the 2+1 dimensional SU(2) gauge theory with $N_h>1$ adjoint Higgs fields. For simplicity, we will generalize the space group symmetries of the model of Ref.~\onlinecite{SSST19} to O($N_h$). We will also neglect the coupling to Fermi surface excitations here, but address this issue in forthcoming work.  We will begin with a lattice discretization of the action of Ref.~\cite{SSST19}, and integrate out the SU(2) gauge field to obtain the following lattice action for the Higgs fields alone
\beq
\label{S0}
S_0 = - \frac{J}{2N_h} \sum_{\langle i j \rangle} H_{a \ell} (i) H_{a m} (i) H_{b \ell} (j) H_{b m } (j) + \frac{u_1}{2 N_h} \sum_i H_{a \ell} (i) H_{a m} (i) H_{b \ell} (i) H_{b m } (i)\,.
\eeq
Here $i$ labels the sites of a cubic lattice, and $H_{a\ell} (i)$ is the real Higgs field, with $a=1,2,3$  the SU(2) adjoint gauge index, and $\ell = 1 \ldots N_h$ the flavor index. Note that $S_0$ is invariant under local SU(2) gauge transformations, but only under global O($N_h$) flavor rotations. We also find it convenient to impose a fixed length constraint on every lattice site, $i$,
\beq
\label{constraint}
\sum_{a \ell} H_{a \ell}^2 (i) = N_h\ .
\eeq
The action $S_0$ comprises a gauge invariant hopping term $J$ that is quartic in Higgs fields, as well as a quartic potential $u_1$ inherited from the original model. 

We now define a gauge-invariant order parameter which is a second-rank traceless tensor in the global O($N_h$) symmetry
\beq
\label{Qlm}
Q_{\ell m} (i) = H_{a \ell} (i) H_{a m}(i) 
- \frac{\delta_{\ell m}}{N_h} H_{a n} (i) H_{a n} (i)
\eeq
This order parameter will diagnose the broken symmetries across the phase diagram. The $\mathbb{Z}_2$ topological order is more subtle to extract directly: we provide evidence for it in the context of the large $N_h$ expansion of $S_0$, and the pattern of symmetry breaking in the Monte Carlo study.

We will study the effective lattice action $S_0$ using both a large $N_h$ saddle point analysis and numerical Monte Carlo (MC) simulations. We will establish that the competition between the two terms in $S_0$ in Eq.~(\ref{S0}) allows for the 3 phases discussed above:\\
({\it i\/}) {\bf Confining:} The Higgs field is fully `disordered' and the global O($N_h$) symmetry is preserved.
This corresponds to the overdoped Fermi liquid in the cuprates.\\
({\it ii\/}) {\bf Trivial Higgs:} The Higgs condensate breaks the SU(2) gauge symmetry down to U(1), which ultimately confines. The O($N_h$) symmetry is broken down to O($N_h -1$).
This is a possible pseudogap phase for the cuprates, and is separated from the confining phase above by a phase transition because of the broken symmetry.\\
({\it iii\/}) {\bf Topological Higgs:} The Higgs condensate breaks the SU(2) gauge symmetry down to $\mathbb{Z}_2$, and there is $\mathbb{Z}_2$ topological order. For $N_h > 3$, the global O($N_h$) symmetry is broken to O(3)$\times$O($N_h - 3)$. This is also a possible pseudogap phase.

The reader will notice that for the special case of $N_h=4$ of interest to us, the patterns of symmetry breaking in the trivial and topological Higgs phases are the same: O(4) is broken down to O(3) in both cases. Nevertheless, as we shall show, it is possible to distinguish these cases by more carefully studying the manner in which O(4) breaks down to O(3). Also, for the cases of $N_h = 2,3$, the topological Higgs phase has no symmetry breaking; nevertheless the topological Higgs phase remains distinct from the confining phase because of its $\mathbb{Z}_2$ topological order.

The outline of the paper is as follows: Section \ref{S:expansion} details the strong-gauge coupling expansion employed to obtain the lattice action for the Higgs field alone $S_0$ (\ref{S0}). In section \ref{S:largeN} we rewrite the effective action $S_0$ using Hubbard-Stratonavich decoupling fields, and subsequently solve the saddle point equations in the limit of $N_h\to\infty$. In this large $N_h$ we produce the phase diagram of the model, which hosts the confined phase, as well as the trivial and topological Higgs phases. In section \ref{S:MC} we turn to a numerical monte Carlo analysis of the effective action $S_0$ (\ref{S0}), with the physically relevant $N_h=4$. We employ two observables to diagnose the various phases. Finally, we discuss our results in section \ref{S:discussion}.

\section{Strong-Coupling Expansion}\label{S:expansion}
We sketch the details of the strong coupling expansion, which also allows us to review the model studied originally \cite{SSST19}. We consider a theory of real Higgs fields $H_{a\ell}$, where $a=1,2,3$ is the SU(2) adjoint gauge index, while $\ell = 1 \ldots N_h$ is the flavor index. 
We will arrive at a theory for this Higgs field which is a discrete time analog of the Schwinger boson theory of antiferromagnets.

\subsection{Lattice Model}
The strong-gauge coupling expansion demands that we work on the lattice. 
The lattice form of the Euclidean Action/Lagrangian is (see e.g. \cite{Laine1995})
\begin{align}
\notag S&=a^3\sum_i\Big\{\left(3\kappa+s\right)\Tr\left[\hat{H}_{m}(i)\hat{H}_{m}(i)\right]-\kappa\sum_{\mu}\Tr\left[\hat{H}_{m}(i) \hat{U}_\mu(i)\hat{H}_{m}(i+a\hat{\bm e}_{\mu}) \hat{U}^\dag_\mu(i)\right] + u_0 \left(H_{am}(i)H_{am}(i)\right)^2 \\
& +u_1 \left(H_{al}(i)H_{am}(i)H_{bl}(i)H_{bm}(i) - \frac{1}{N_h}(H_{am}(i)H_{am}(i))^2\right)+ \beta\sum_{\mu>\nu} \left[1-\frac{1}{2}\Tr \ \hat{G}_{\mu\nu}(i) \right]\Big\}
%V(\hat{H}_m)&=u_0 \left(H_{am}H_{am}\right)^2+u_1 \left(H_{al}H_{am}H_{bl}H_{bm} - \frac{1}{N_h}(H_{am}H_{am})^2\right)
\end{align}
where $\kappa= 4/a^2$,  $\beta= 4/(g a^2)^2$ and $a$ is the lattice spacing; summation is over the elementary unit cell, whereby $\hat{\bm e}_{\mu}=\{\hat{e}_x, \hat{e}_y, \hat{e}_\tau\}$; trace is over gauge indices, and summation over flavours $m$ is implied. The Higgs field, gauge field link, and Yang-Mills plaquette operators are given by
\begin{align}
\hat{H}_m(i)&=H_{am}(i)\tau^a\\
\hat{U}_\mu(i)&=e^{i a A_{a\mu}(i)\tau^a}\\
\hat{G}_{\mu\nu}(i)&=\hat{U}_\mu(i)\hat{U}_\nu(i+a\hat{\bm e}_{\mu})\hat{U}^\dag_\mu(i+a\hat{\bm e}_{\mu}+a\hat{\bm e}_{\nu})\hat{U}^\dag_\nu(i+a\hat{\bm e}_{\nu})\label{fields}
\end{align}
where $\tau^a$ are Pauli matrices, with normalization $\Tr\left[\tau^a\tau^b\right]=\delta^{ab}/2$. The gauge link and plaquette operators follow the usual lattice-gauge transformation laws \cite{Laine1995}. From \equref{fields} we see that the Yang-Mills term $\hat{G}_{\mu\nu}(i)$ is just the parallel transport around the elementary unit cell.

\subsection{Strong-Coupling Expansion}
\begin{figure}[t]
 \includegraphics[width=160mm]{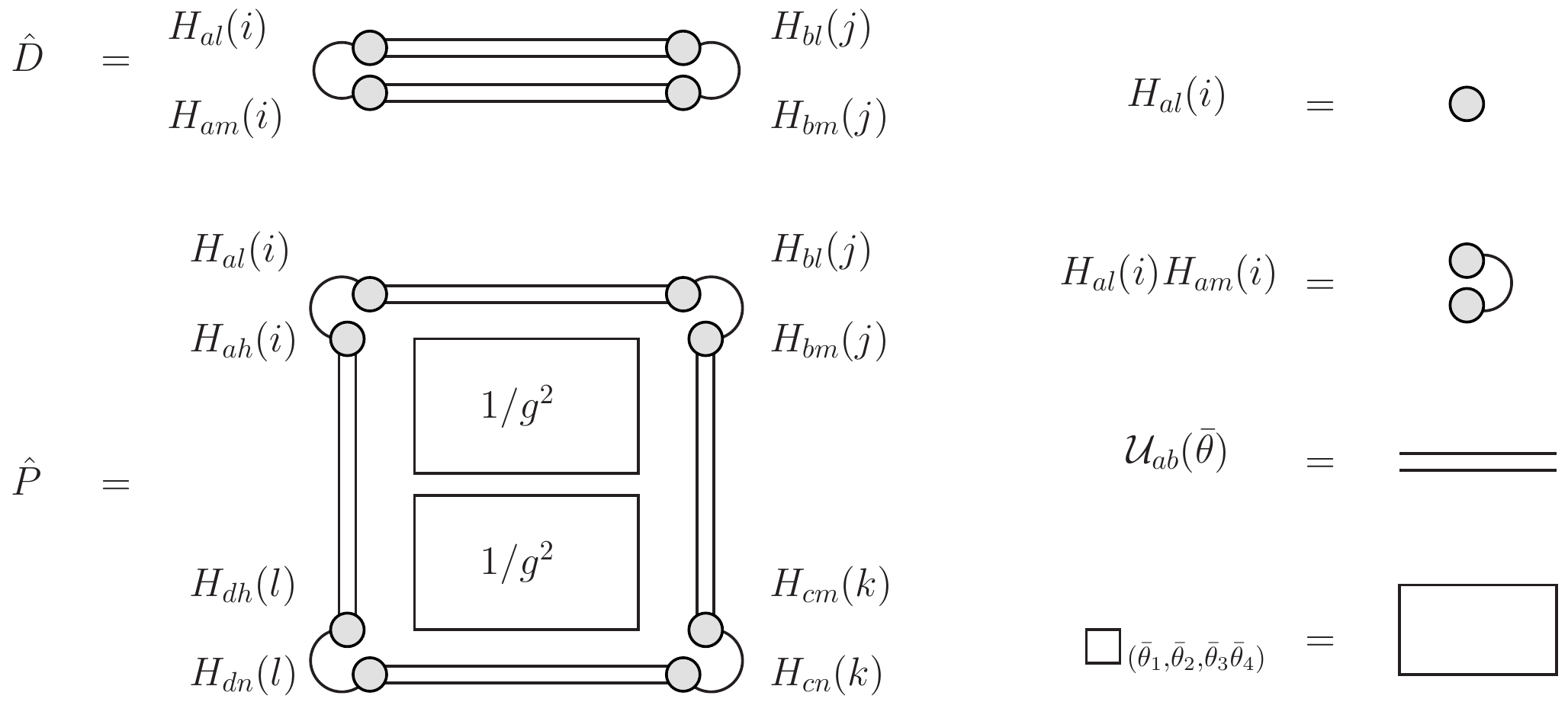}
\caption{{Diagrammatic representation of the strong-gauge coupling expansion. Operators $\hat{D}, \hat{P}$ correspond to the double Higgs and Higgs plaquette. We see that $\hat{D}$ does not require any gauge plaquette terms $\square$ at leading order, while the $\hat{P}$ requires two gauge plaquettes at leading order. Definitions are shown on the right hand side.} \label{f:diagrams}}
\end{figure}

Due to strong coupling $g\to\infty$, the kinetic Yang-Mills action is neglected and then each gauge link, $U_\mu(i)$, is an independent random $SU(2)$ matrix.  We choose to parameterise each such link by the three Euler angles $\bar\theta=\{\theta,\psi,\phi\}$,
\begin{align}
U(\bar\theta)=\cos\theta\ \hat\sigma_0+i\sin\theta\sin\psi\cos\phi \ \hat\sigma_1+ i\sin\theta\cos\psi \ \hat\sigma_2 + i\sin\theta\sin\psi\sin\phi \ \hat\sigma_3\ .
\end{align}
At strong-coupling we may treat the random Higgs-hopping term $\Tr\left(\hat{H}_{m}(i) \hat{U}_\mu(i)\hat{H}_{m}(i+\hat{\mu}) \hat{U}^\dag_\mu(i)\right)$ as a perturbation, even though this is not formally an expansion in $1/g^2$. Expanding the partition function in this hopping term generates terms such as,
\begin{align}
%{\cal Z}_0^{-1}\int [DU][DH] \Tr\left(\sigma_a U({\bar\theta}) \sigma_b U^\dag({\bar\theta})\right) e^{i{\cal S}_0[H]} \equiv \braket{\Tr\left(\sigma_a U({\bar\theta}) \sigma_b U^\dag({\bar\theta})\right)}_{{\cal Z}_0}&=0.
\int [DU] \Tr\left(\sigma_a U({\bar\theta}) \sigma_b U^\dag({\bar\theta})\right) \equiv \braket{\Tr\left(\sigma_a U({\bar\theta}) \sigma_b U^\dag({\bar\theta})\right)}_{U}&=0.
\end{align}
The expectation value must vanish since it transforms nontrivially under $SU(2)$ transformations; the integration over $[DU]$ evaluates to zero. 
To consider higher-order terms, it is convenient to first define the adjoint matrix
\begin{align}
{\cal U}_{ab}(\bar\theta)&\equiv\frac{1}{4}\Tr\left(\sigma_a U(\bar\theta) \sigma_b U^\dag(\bar\theta)\right), 
\end{align}
with normalization $\Tr\left(\sigma_a\sigma_a\right)=6$. The non-vanishing terms in the expansion of the Higgs hopping will need to be invariant in the adjoint indices. We find that for example (with no contraction over $a,b$ indices),
\begin{align}
\label{HHexp}
 \braket{{\cal U}_{ab}(\bar\theta)}_U&=0\\
 \braket{{\cal U}_{ab}(\bar\theta){\cal U}_{ab}(\bar\theta)}_U&=\frac{1}{12}\\
  \braket{{\cal U}_{ab}(\bar\theta_1){\cal U}_{ab}(\bar\theta_2)}_U&=0.
 \end{align}
Here $\bar\theta_1\neq\bar\theta_2$ signifies different gauge links. The nonzero expectation value above implies that the lowest order expansion does not require a $1/g^2$ gauge-plaquette expansion to compensate, see Figure \ref{f:diagrams} for a diagrammatic representation. We call this term the double Higgs link $\hat{D}$. The contribution to the action is then the gauge field averaged
\begin{align}
\label{Dlink}
\braket{\hat{D}}_U\sim \kappa^2\sum_{\braket{i j}} H_{al}(i)H_{am}(i)H_{bl}(j)H_{bm}(j) ,
\end{align}
which is manifestly gauge-invariant. Higher order terms are derived in the Appendix, e.g. the Higgs-plaquette term $\hat{P}$ of Figure \ref{f:diagrams}. We neglect such a term in the present analysis, since we will find that the double Higgs link, $\hat{D}$, is already sufficient to generate the expected topological properties of the underlying gauge theory. 

%\subsection{Effective Action}

Imposing the constraint in Eq.~(\ref{constraint}), and re-exponentiating the double Higgs link term (\ref{Dlink}), we arrive at the effective action
$S_0$ in Eq.~(\ref{S0}) on the three-dimensional cubic lattice. 

\section{Large $N_h$ limit}\label{S:largeN}

We set up the large $N_h$ expansion by writing the partition function as
\bea
Z &=& \int \prod_{\langle i j \rangle} d A_{ab} (i,j) \prod_{i} d B_{ab} (i) \prod_i d \lambda (i) \prod_i d H_{a\ell} (i)\, e^{-S} \nn
S&=& \sum_{\langle ij \rangle} \left[ \frac{N_h \left[A_{ab}(i,j) \right]^2}{2J} - A_{ab} (i,j) H_{a \ell} (i) H_{b \ell} (j) \right] \nonumber \\
&~&+ \sum_i \left[ \frac{N_h \left[B_{ab}(i) \right]^2}{8u_1} + i \frac{B_{ab} (i)}{2} H_{a \ell} (i) H_{b \ell} (i) + i \frac{\lambda (i)}{2} \left( H_{a \ell} (i) H_{a \ell} (i) - N_h \right) \right] \label{Z1}
\eea 
For the fluctuations to be stable, the signs and factors of $i$ have been chosen assuming $u_1 > 0$. But the formalism works for both signs of $u_1$, and we just have to rotate the contour for $B$ in the fluctuations for $u_1 < 0$. We are interested in the case of $J>0$.

\subsection{Saddle-Point Phase Diagram}
\begin{figure}[t]
\includegraphics[width=140mm]{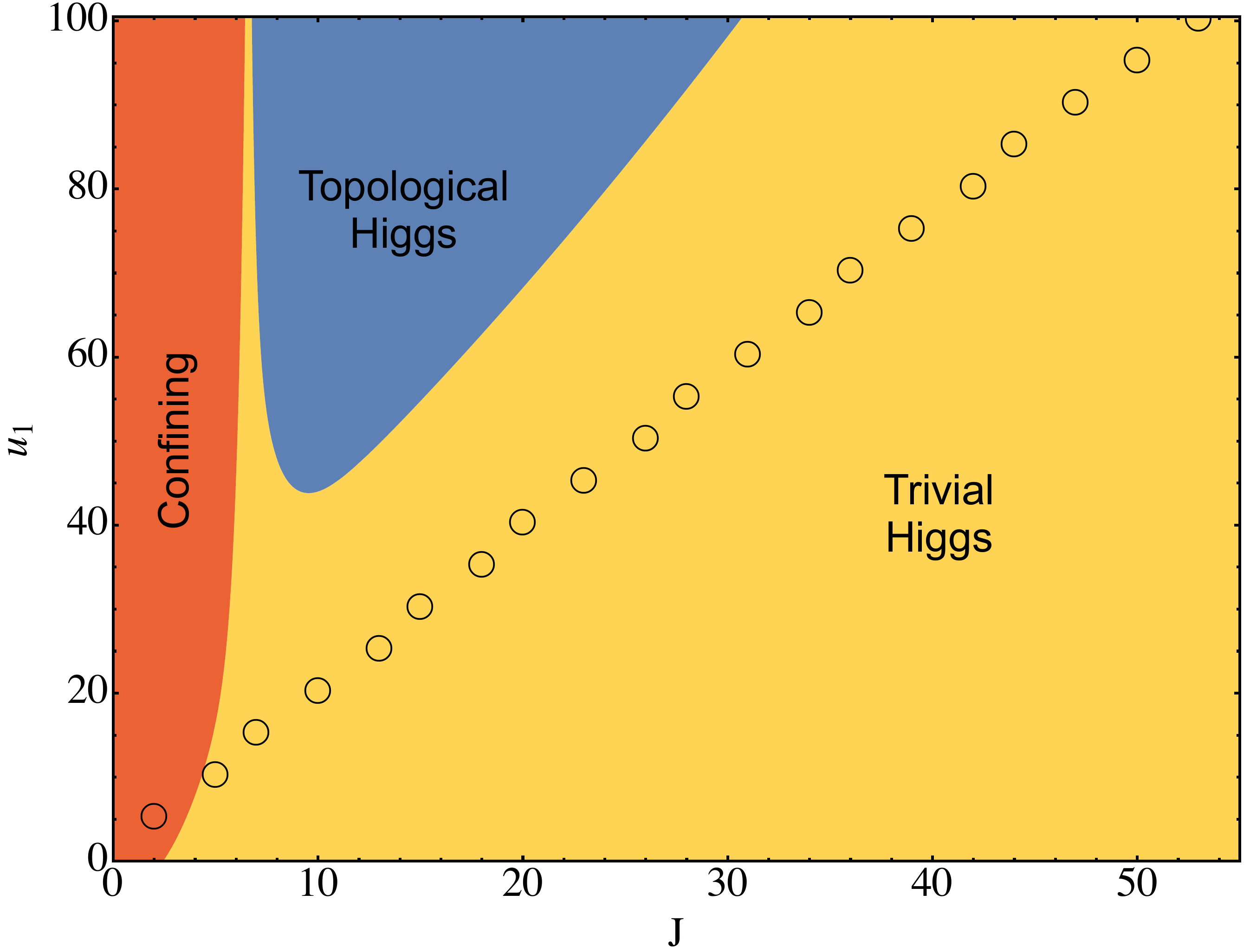}
\caption{{Phase Diagram. Coloured regions correspond to the phases found in the saddle point approximation ($N_h\to\infty$). Black empty circles correspond to the topological-to-trivial phase transition as found in MC simulations of the parent model at $N_h=4$, and with system size $L^3$;  $L=12$.} \label{f:phase}}
\end{figure}

We begin by providing the results of the saddle point analysis -- the details of which are left for subsections \ref{S:disordered} and \ref{S:ordered}. 
Comparing the free energies of the disordered, topological and trivial phases obtained in the saddle point analysis we arrive at the phase diagram shown in Figure \ref{f:phase}.  Noteworthily, we find that all phase boundaries are of first order. Also shown in Figure \ref{f:phase} is the topological-to-trivial phase transition as determined by MC simulations of the parent action $S_0$ (\ref{S0}), for which we take the physical number of Higgs flavours $N_h=4$. Details of the identification of the phase transition from MC simulations are provided Section \ref{S:MC}.

We now outline how the saddle point solutions were obtained, further details are provided in Appendix \ref{A:saddle}.

\subsection{Confining Phase} \label{S:disordered} 
In the confining phase, the Higgs field is fully disordered and maintains the $O(N_h)$ global symmetry. This places no restrictions on the other decoupling fields appearing in action $S$ (\ref{Z1}), instead we will assume a gauge-invariant saddle point of $S$ (\ref{Z1}). In the limit $N_h \rightarrow \infty$ limit, the saddle point fields are then,
\begin{subequations}\bea
H_{al}(i)&=&0 \\
A_{ab} (i,j) &=& \delta_{ab} A_0 \\
iB_{ab} (i) &=& \delta_{ab} B_0 \\
i \lambda (i) &=& \overline{\lambda}
\eea \end{subequations}
It follows that the propagator of the Higgs field is diagonal in flavor and color indices and is given by
\beq
G(k) =  \frac{1}{A_0 (6 - 2\cos(k_x) - 2\cos (k_y) - 2\cos( k_\tau) ) + m^2}
\eeq
where the mass gap relates to the saddle point fields via
\beq m^2 = \overline{\lambda} + B_0 - 6 A_0. \eeq
In the large $N_h$ limit, the free energy density, $F$, obtained by integrating over the Higgs fields $H_{al}(i)$ is
\beq
\frac{F}{N_h} = 3 \left[\frac{3 A_{0}^2}{2 J}  - \frac{B_{0}^2}{8 u_1} \right]
- \frac{\overline{\lambda}}{2}  - \frac{3}{2} \int_{-\pi}^{\pi} \frac{d^3 k}{8 \pi^3} \ln \Bigl[ G (k) \Bigr], \label{f0}
\eeq
which relates to the partition function via $Z = e^{-FV}$, where $V$ is the Euclidean volume. 
Minimising the free energy, the saddle point equations determining $A_0$, $B_0$, and $m^2$ are obtained
\begin{subequations}\bea
3 \int_{-\pi}^{\pi} \frac{d^3k}{8 \pi^3} G(k) &=& 1, \label{soa} \\ 
J \int_{-\pi}^{\pi} \frac{d^3k}{8 \pi^3} \cos (k_x) G(k) &=& A_0 , \label{sob} \\
B_0 &=& \frac{2}{3} u_1 .
\eea  \label{s0} \end{subequations} 

There are two classes of solutions to these saddle point equations in disordered phase: those with $A_0=0$, and those with $A_0\neq0$. In the first case, the saddle point admits a particularly simple solution, 
\begin{align}
A_0=0; \ \ \ \bar\lambda=3-B_0=3-\frac{2u_1}{3}.
\end{align}
From which it follows that the free energy is independent of $J$,
\begin{align}
\frac{F}{N_h}&=\frac{u_1}{6}-\frac{3}{2}(1-\ln3).
\end{align}
Details presented in the appendix show that the $A_0\neq0$ solutions always posses a higher free energy in the $(J, u_1)$ phase diagram, and hence would only appear as metastable states. 

\subsection{Higgs phases}\label{S:ordered} 

In the ordered phases, we proceed as in Ref.~\cite{DPSS}. Moreover, we follow Ref.~\cite{SSST19}, and note that --
by the singular value decomposition theorem -- any Higgs field can be written in the form
\bea
\label{SVD}
H_{al} &=& O_{1;ab}W_{bm}O_{2;ml}
\eea
where $O_1$ and $O_2$ are orthogonal matrices in color and flavor spaces respectively, and $W$ is a
rectangular matrix with only $p\equiv \min(3, N_h)$ non-zero elements along its diagonal, which are all non-negative. 
Owing to this decomposition, we write the Higgs field using the following ansatz, 
\beq
H_{a\ell}(i) = \sqrt{N_h} H_{0a} \delta_{a\ell} + H_{1a \ell}(i)
\eeq
where $H_{0a}$ is a possible non-zero, site-independent saddle-point value, and we integrate over the additional fluctuations, $H_{1al}(i)$, around the saddle point.
We allow the other saddle-point variables to depend upon the color indices by writing,
\bea 
A_{ab} (i,j) &=& \delta_{ab} A_{0a} \nn
iB_{ab} (i) &=& \delta_{ab} B_{0a} \nn
i \lambda (i) & = & \overline{\lambda} \,.
\eea 
In the large $N_h$ limit, the free energy density, $F$, obtained by integrating over the $H_{1al}(i)$ is %(we drop certain terms that do not contribute at large $N_h$)
\bea
\frac{F}{N_h} &=& \sum_a \left[\frac{3 A_{0a}^2}{2 J} - 3 A_{0a} H_{0a}^2 - \frac{B_{0a}^2}{8 u_1} + \frac{B_{0a}}{2} H_{0a}^2 \right]
+ \frac{\overline{\lambda}}{2} \left( \sum_a H_{0a}^2 - 1 \right) \nn
&~& - \frac{1}{2} \sum_a \int_{-\pi}^{\pi} \frac{d^3 k}{8 \pi^3} \ln \Bigl[ G_a (k) \Bigr] \label{f1}
\eea
where the Greens function obtains a color index, and is given by,
\beq
G_a (k) = \frac{1}{ \overline{\lambda} + B_{0a} - 2 A_{0a} \left(\cos(k_x) + \cos(k_y) + \cos(k_\tau) \right) }.
\eeq
We now study the saddle point equations of (\ref{f1}) with respect to $H_{0a}$, $A_{0a}$, $B_{0a}$, and $\overline{\lambda}$. (Note: we cannot just globally minimize $F$ because of the $i$'s in (\ref{Z1}).) The saddle point of the action with respect to the $H_{0a}$ gives us the three equations
\begin{subequations}\beq
\overline{\lambda} H_{0a} = (- B_{0a} + 6 A_{0a}) H_{0a} ~,~\mbox{for all $a$, with no sum over $a$.}
\label{s1}
\eeq
We do not cancel out the $H_{0a}$ in (\ref{s1}) because $H_{0a}$ could vanish for some $a$.
The saddle point with respect to $\overline{\lambda}$ is
\beq
\sum_a \left[ H_{0a}^2 + \int_{-\pi}^{\pi} \frac{d^3k}{8 \pi^3} G_a(k) \right] = 1\,. \label{s2}
\eeq
Finally, the saddle point equations with respect to $A_{0a}$ and $B_0$ are
\bea
J \int_{-\pi}^{\pi} \frac{d^3k}{8 \pi^3} \cos (k_x) G_a (k) &=& A_{0a} - J H_{0a}^2 \\
B_{0a} &=& 2 u_1 \left[ H_{0a}^2 + \int_{-\pi}^{\pi} \frac{d^3k}{8 \pi^3} G_a(k) \right] \label{s3}
\eea \label{SEsOrdered} \end{subequations}
Note that \equref{SEsOrdered} reduces to (\ref{s0}) when $H_{0a} = 0$.

We now have to solve the 10 equations (\ref{s1},\ref{s2},\ref{s3}) for the 10 variables $H_{0a}$, $A_{0a}$, $B_{0a}$, and $\overline{\lambda}$ as a function of $J$ and $u_1$.
There will be two types of solutions: one in which only one of the $H_{0a}$ is non-zero,
and the other in which all $H_{0a}$ are equal to each other -- this corresponds to the topological phase, as deduced by the global and gauge symmetry breaking patterns, which is discussed in \cite{SSST19}, yet we outline the argument here for continuity of presentation: The gauge symmetry is $SU(2)$, condensing one Higgs flavour reduces this to a remnant $U(1)$ which corresponds to rotations about the axis set by the condensed field, while all Goldstone modes are Higgsed (i.e. gapped). It is well established that the gapped $U(1)$ gauge theory is ultimately in a confining phase, yet the confinement length scale depends on the details of the system. This is the trivial Higgs phase, and is achieved in the saddle point by just one $H_{al}\neq0$. We mention that Berry phase interference effects could act to deconfine the $U(1)$ gauge theory \cite{SenthilDQC}; we do not consider such effects in this work. Alternatively, condensing multiple Higgs flavours, with some orthogonal components, breaks the $SU(2)$ gauge down to $\mathbb{Z}_2$ (since the Higgs fields themselves are in the adjoint representation). This remnant $\mathbb{Z}_2$ gauge theory is naturally deconfined, supporting $\mathbb{Z}_2$ topological order. Condensing multiple Higgs flavours is achieved by the saddle point with all $H_{0a}\neq0$ and equal to each other.    

The true ground state configuration will be the saddle point solution for which the free energy (\ref{f1}) is minimised. We will now compute the saddle point equations and free energy for both cases. 

\subsection{Topological Higgs Solutions}
The topological solution can be obtained analytically. In this phase three classes of solutions arise, here we will present just the dominant one, the other two are left for the appendix. 
For this solution we  have $H_{01}=H_{02}=H_{03}\equiv H$. By inspection of the saddle point equations, the solution has $A_{01}=A_{02}=A_{03}\equiv A$ and $B_{01}=B_{02}=B_{03}\equiv B$. The solutions are (with $\sigma=\pm$)
\begin{align}
\label{Atopo}
A_{\sigma}&=\frac{J}{6}+\sigma\frac{1}{6}\sqrt{J^2-6J}\\
\label{Btopo}
B&=\frac{2}{3}u_1\\
\bar{\lambda}_\sigma&=6A_\sigma - B\\
H^2&=\frac{A_\sigma}{J}-\frac{\gamma_2}{A_\sigma} \ \ \ \ \text{($H$ is independent of the $\sigma$ index)}
\end{align}
where the constant $\gamma_2$ (and for later use $\gamma_1$) are defined as 
\begin{align}
\gamma_1&=\int_{-\pi}^\pi\frac{d^3k}{(2\pi)^3}\frac{1}{6-2\sum_\mu\cos k_\mu}, \ \ \ \gamma_2=\int_{-\pi}^\pi\frac{d^3k}{(2\pi)^3}\frac{\cos k_x}{6-2\sum_\mu\cos k_\mu}, \ \ \ 
\gamma_1-\gamma_2=\frac{1}{6}. 
\end{align}

The free energy in this phase can be written solely in terms of $A$ and $B$, we find
\begin{align}
\frac{F}{N_h}&=\frac{9 A^2}{2J}-\frac{3B^2}{8u_1}-\frac{1}{2}(6A-B)+\frac{3}{2}\int_{-\pi}^\pi\frac{d^3k}{(2\pi)^3}\ln\left(A(6-2\sum_\mu\cos k_\mu)\right).
\end{align}
This is straightforward to evaluate using the relations above (\ref{Atopo}) and (\ref{Btopo}). Using notation $A_\sigma$ with $\sigma=\pm$, we get the expression,
\begin{align}
\frac{1}{N_h}F(u_1, J)&=\frac{1}{24} \left(-6 J-6 \sigma  \left(\sqrt{(J-6) J}+3 \sigma \right)-36 \ln
   \left(\frac{6}{\sqrt{(J-6) J} \sigma +J}\right)+4 u_1\right) + c\\
c&\equiv -\frac{3}{2}\int_{-\pi}^\pi\frac{d^3k}{(2\pi)^3}\ln\left((6-2\sum_\mu\cos k_\mu)\right)=-2.51008.
 \end{align}
 The $\sigma=+1$ root minimises this free energy.
We see the simple result that $F(u_1,J)$ is linear in $u_1$ (for the topological solution), moreover the coefficient $1/6$ is the same as the disordered phase, hence the critical point separating these two phases is independent of $u_1$ -- although the direct transition between disordered and topological phases is masked by the trivial phase, as shown next. %Implying that $u_1<0$ has a lower energy than $u_1>0$. On the other hand we expect that $u_1>0$ gives a topological solution and $u_1<0$ gives a trivial solution. So it must be the case that for $u_1<0$ the trivial solution has a lower free energy. 
 %In Figure \ref{f:FE} we show the corresponding free energy for $u_1=\pm 1$.

\subsection{Trivial Higgs Solution}
The trivial solution is more difficult. In this phase we set $H_{01}\equiv H$ and $H_{02}=H_{03}=0$. Once again there are multiple classes of solutions, and we present just the dominant -- leaving the other for the appendix.
By inspection, we can set $A_{01}\equiv A_1$ and $A_{02}=A_{03}\equiv A_2$ and $B_{01}\equiv B_1$ and $B_{02}=B_{03}\equiv B_2$. We can massage the saddle point expressions analytically to express all fields in terms of just $A_1$, 
\begin{align}
\label{B1}
B_1&=-\bar\lambda + 6A_1\\
\label{B2}
B_2&=u_1+\frac{1}{2}(\bar\lambda-6A_1)\\
\label{H}
H^2&=\frac{A_1}{J}-\frac{\gamma_2}{A_1} \\
\label{lambda}
\bar\lambda&=6A_1-\frac{2u_1}{J}A_1-\frac{u_1}{3 A_1}.
\end{align}
Finally, we have reduced the saddle point equations to a self-consistent equation in the single field variable $A_1$, which reads
\begin{align}
\label{self}
A_2(A_1)&=J\int_{-\pi}^\pi\frac{d^3k}{(2\pi)^3}\frac{\cos k_x}{W(A_1)-2A_{2}(A_1)\sum_\mu\cos k_\mu}.
\end{align}
We notice one simple analytic solution: $A_2=0$. Setting $A_2=0$, we get from \equref{s3} a single polynomial equation in a single variable $A_1$ 
\begin{align}
1=H(A_1)^2+\frac{\gamma_1}{A_1} + \frac{2}{\bar\lambda(A_1)+B_2(A_1)}
\end{align}
which gives four roots: denoted $A_1^{(i)}, \ i=1,2,3,4$. The roots can be obtained analytically, although the expressions are lengthy.

\section{Monte Carlo Results}\label{S:MC}
We perform MC simulations of the parent action $S_0$ (\ref{S0}), with the physical value of $N_h=4$. Details of the MC updates schemes are provided in Appendix \ref{MC details}. Here we consider two diagnostics of the phases and transitions:\\  
(i) The first diagnostic is the eigenvalues $\omega_i >0$, with $i=1,2,3$ from the singular value decomposition of the Higgs field $H_{0a}$ (\ref{SVD}). The saddle point analysis predicts that the trivial phase will posses inequivalent eigenvalues, whereby $\omega_1 > 0$ and $\omega_2=\omega_3=0$. Meanwhile the topological phase will have three degenerate non-zero eigenvalues, $\omega_1=\omega_2=\omega_3 > 0$.\\ 
(ii) The second diagnostic is the scalar observable
\begin{align}
\Phi=\frac{1}{V}\sum_{l,m}\left(\sum_iQ_{lm}(i)\right)^2
\end{align} 
where $Q_{lm}(i)$ is the gauge invariant order parmeter (\ref{Qlm}). According to the saddle point analysis, $\Phi$ shows markedly different behaviour as a function of $(u_1, J)$ for the topological and trivially ordered configurations. Moreover from (\ref{SVD}), we see that $Q_{\ell m}$ is related by global O(4) rotations to a diagonal matrix
\begin{equation}
Q_{\ell m} = O_{2,\ell' \ell} O_{2, m' m} \left( 
\begin{array}{cccc}
\frac{3}{4}\omega_1^2 - \frac{1}{4} \omega_2^2 - \frac{1}{4}\omega_3^2 & 0 & 0 & 0 \\
0 & \frac{3}{4}\omega_2^2 - \frac{1}{4} \omega_3^2 - \frac{1}{4}\omega_1^2 & 0 & 0\\
0 & 0 & \frac{3}{4}\omega_3^2 - \frac{1}{4} \omega_1^2 - \frac{1}{4}\omega_2^2 & 0 \\
0 & 0 & 0 & -\frac{1}{4}\omega_1^2 - \frac{1}{4} \omega_2^2 - \frac{1}{4}\omega_3^2
\end{array}
\right)_{\ell' m'} 
\end{equation}
Note that the diagonal elements equal $\omega_1^2 (3/4,-1/4,-1/4,-1/4)$ in the trivial Higgs phase, and equal $\omega_1^2 (1/4,1/4,1/4,-3/4)$ in the topological Higgs phase. These configurations of $Q_{\ell m}$ are not equivalent to each other, and cannot rotated to each other by a O(4) transformation. 

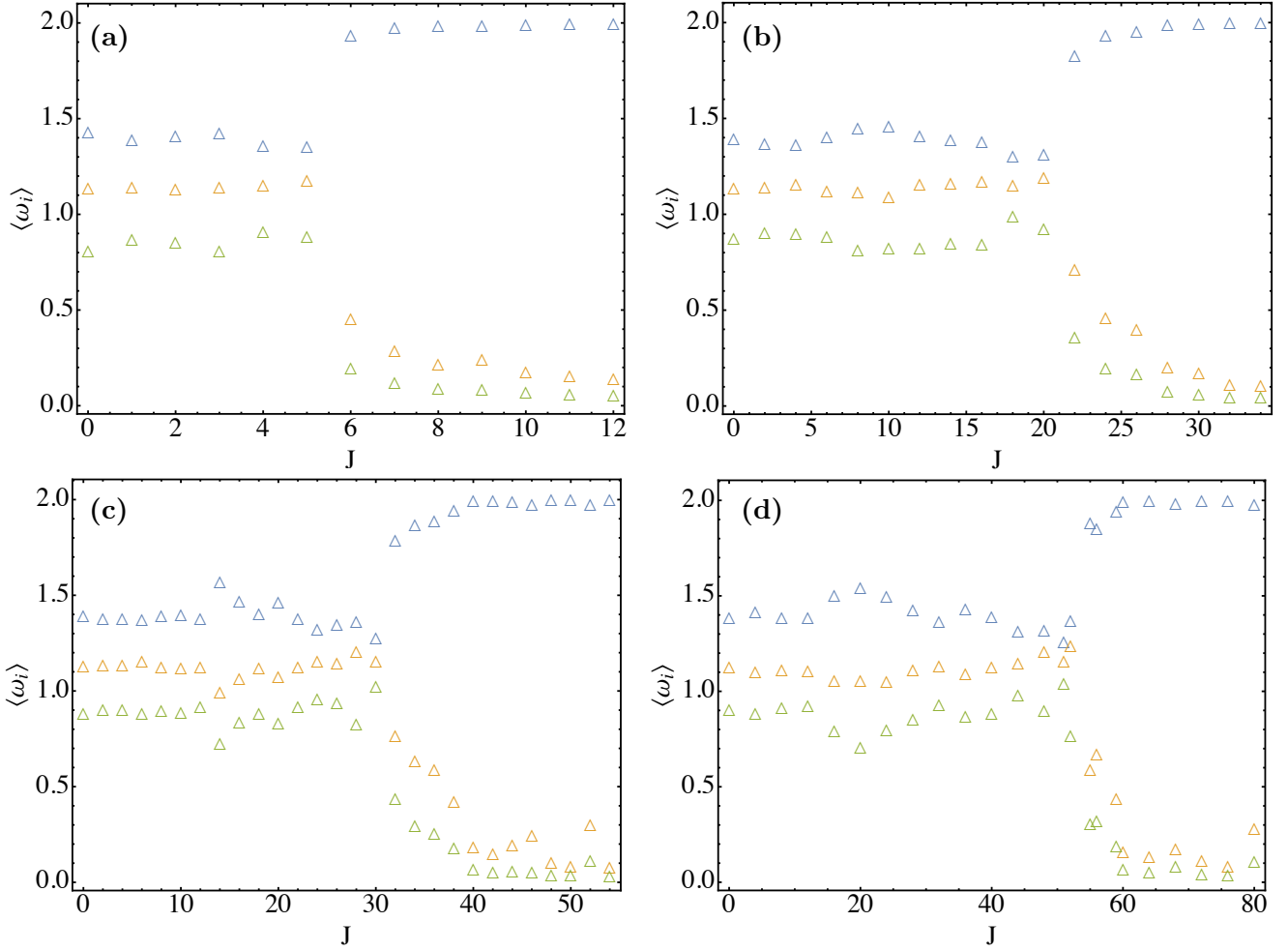
\begin{figure}[t]
\includegraphics[width=85mm]{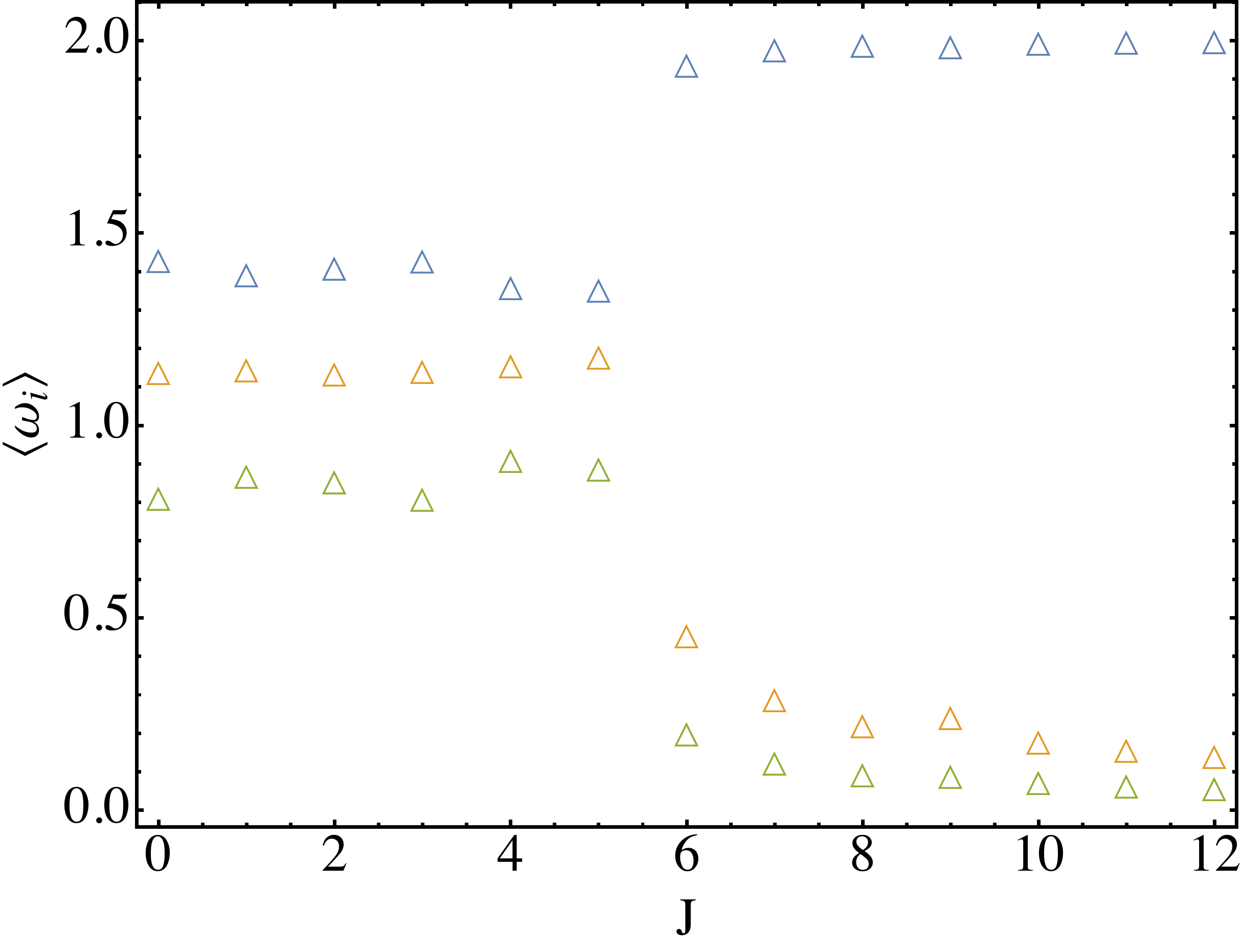}\hspace{0.25cm}
\includegraphics[width=85mm]{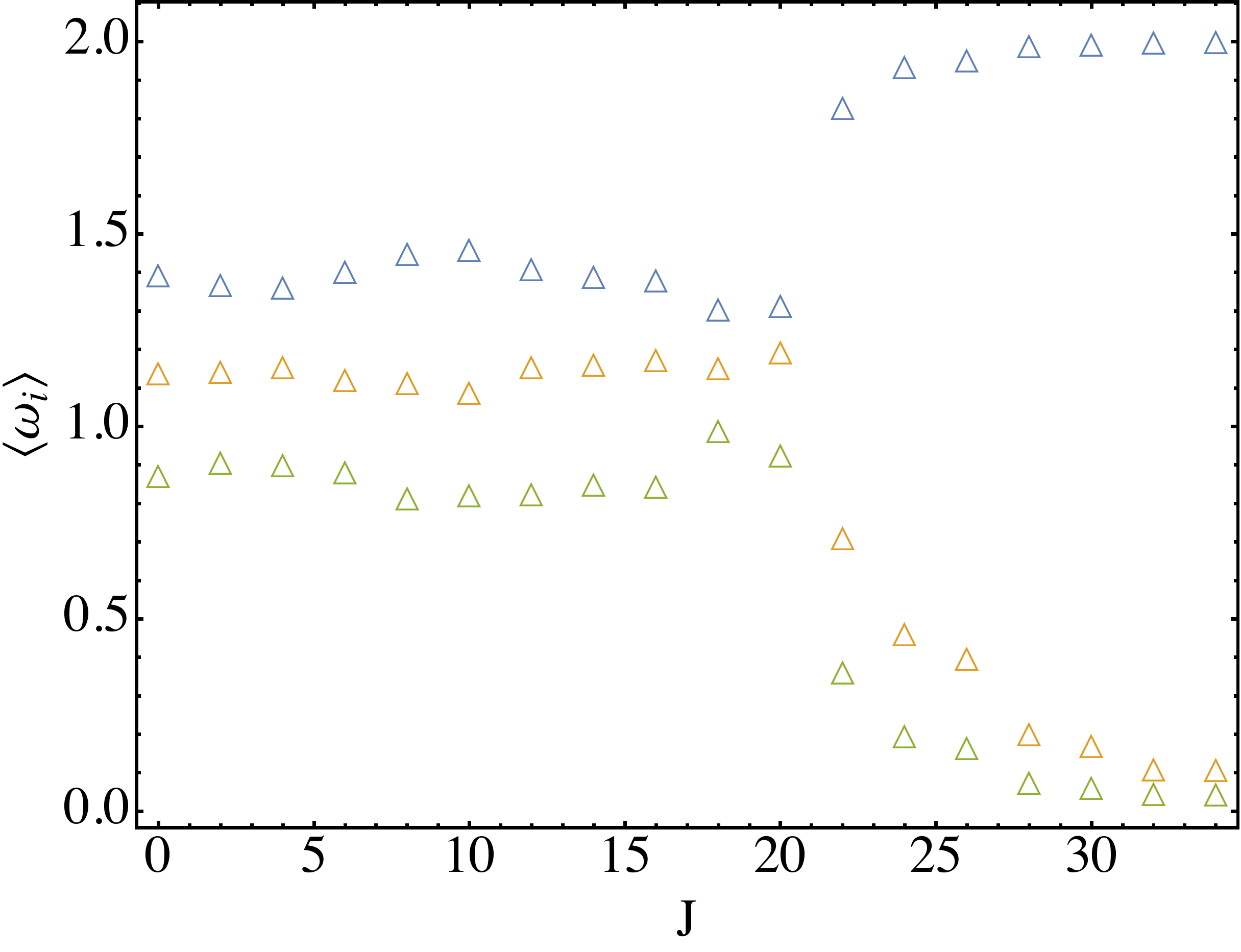}
\includegraphics[width=85mm]{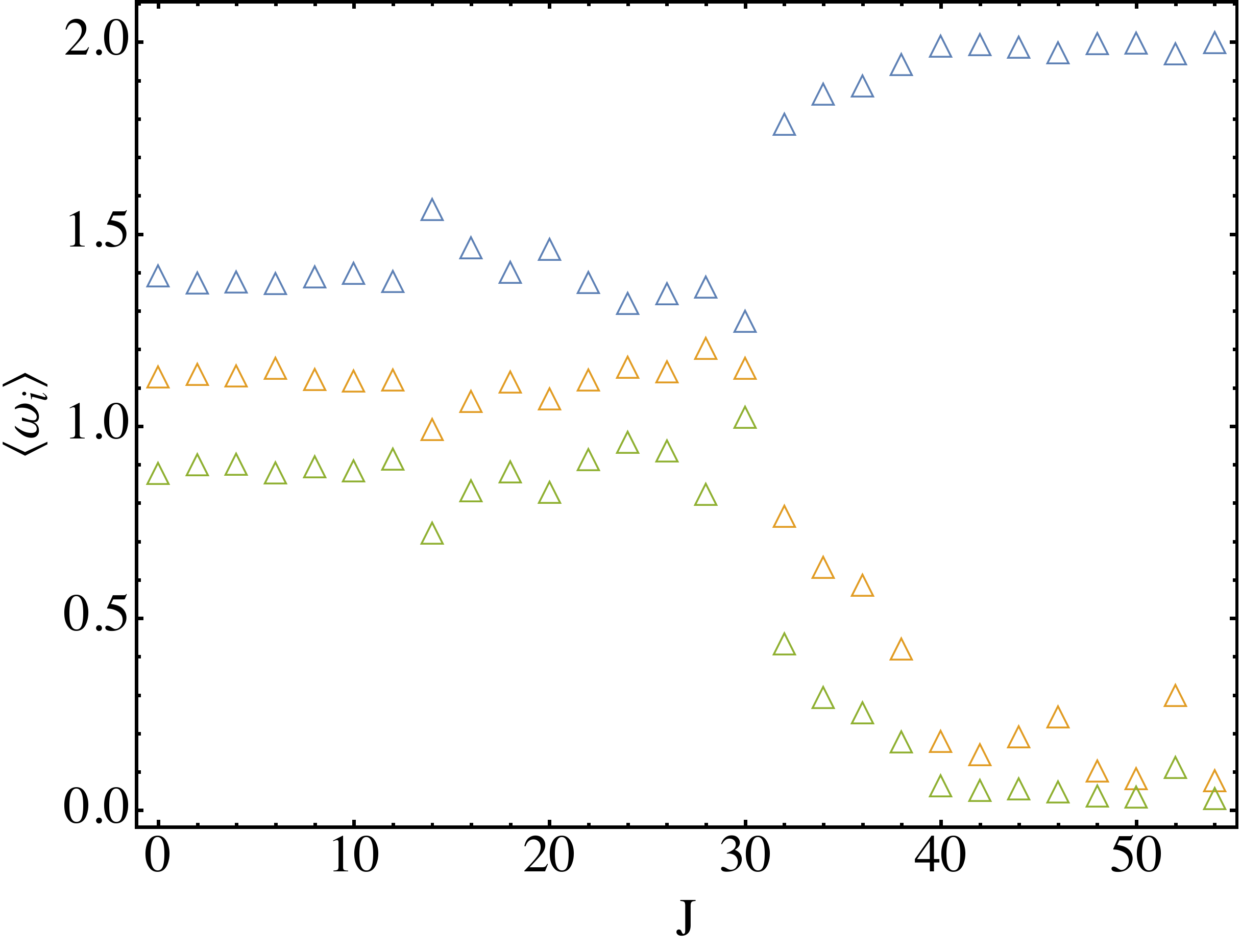}\hspace{0.25cm}
\includegraphics[width=85mm]{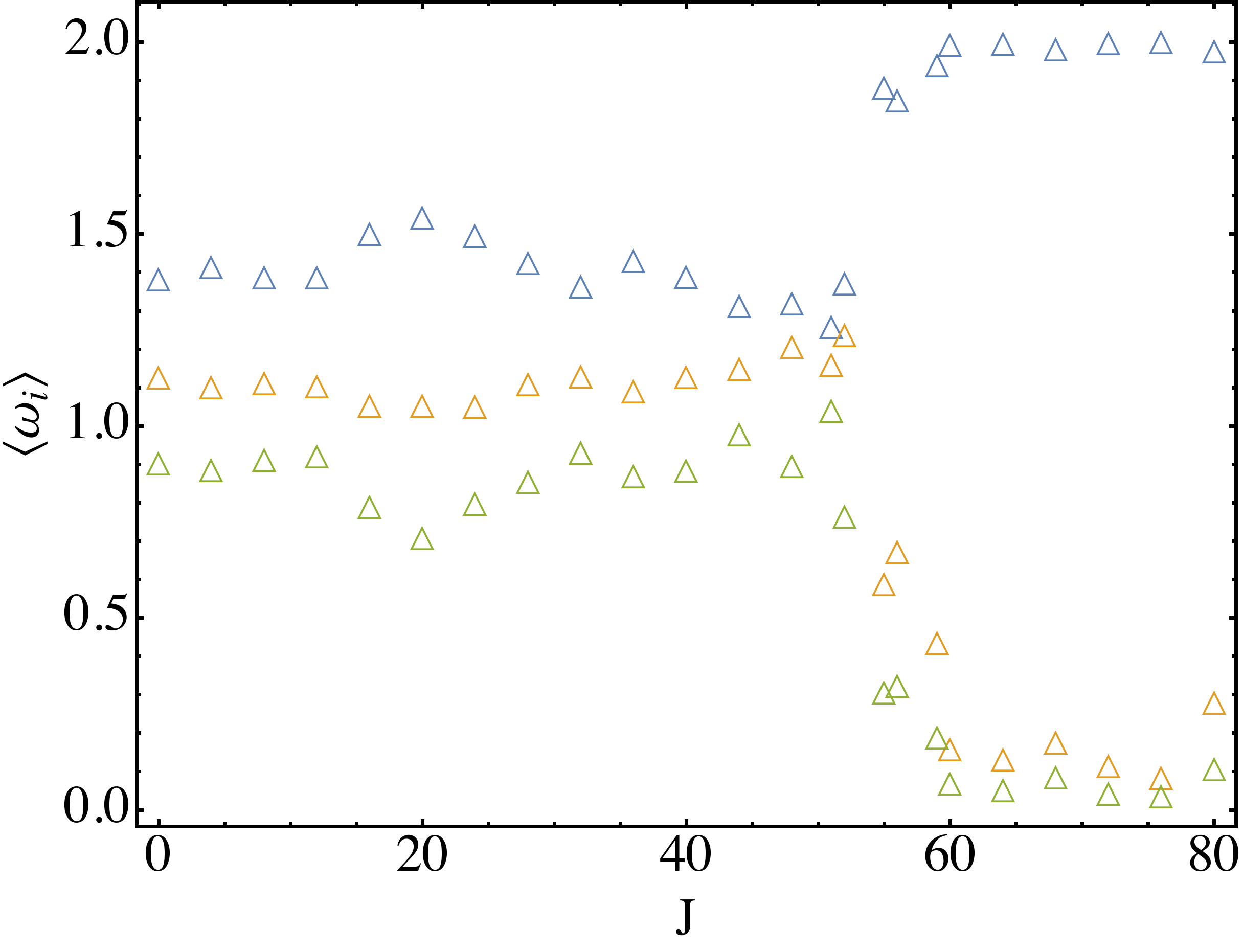}
\begin{picture}(0,0) 
\put(-465,355){\textbf{(a)}}
\put(-210,355){\textbf{(b)}}
\put(-465,170){\textbf{(c)}}
\put(-210,170){\textbf{(d)}}
\end{picture}
\caption{{System averaged onsite eigenvalues $\braket{\omega_i}$ of the Higgs field $H_{al}(i)= O_{1;ab}W_{bm}(i)O_{2;ml}$ obtained via SVD. (a), (b), (c), and (d) correspond to $u_1=\{10, 40 ,60, 100\}$. System size $L=12$.} \label{f:Eval}}
\end{figure}

To obtain the first observable in the MC simulations, after the ground state is reached, we perform the SVD (\ref{SVD}) at each site and average over the system, giving the averaged eigenvalues $\braket{\omega_i}$. In Figure \ref{f:Eval} we plot the averaged eigenvalues $\braket{\omega_i}$, and their evolution with $J$ for various $u_1$. These are obtained from MC simulations on lattices of size $L^3$ with $L=12$. The phase transition between topological and trivial order is identified with the large discontinuity in the eigenvalues as they transition from nearly degenerate, to non-degenerate. The corresponding phase boundary estimate has already been plotted in Figure \ref{f:phase}, from which we see qualitatively the same trend in transitioning from the disordered or topological into the trivial phase, i.e. a linear dependence of $u_{1,c} \propto J_c$. This agreement indicates that $1/N_h$ corrections do not destabilize the topological phase.  

An additional feature is apparent from the eigenvalues for $u_1>10$ and $J<J_c$, see e.g. $J\sim20$ in Figure \ref{f:Eval}(d). This is perhaps a sign of the small window (in $J$) of trivial phase wedged between the disordered and topological phases -- as predicted by the saddle point analysis and shown in Figure \ref{f:phase}.

\begin{figure}[t]
\includegraphics[width=85mm]{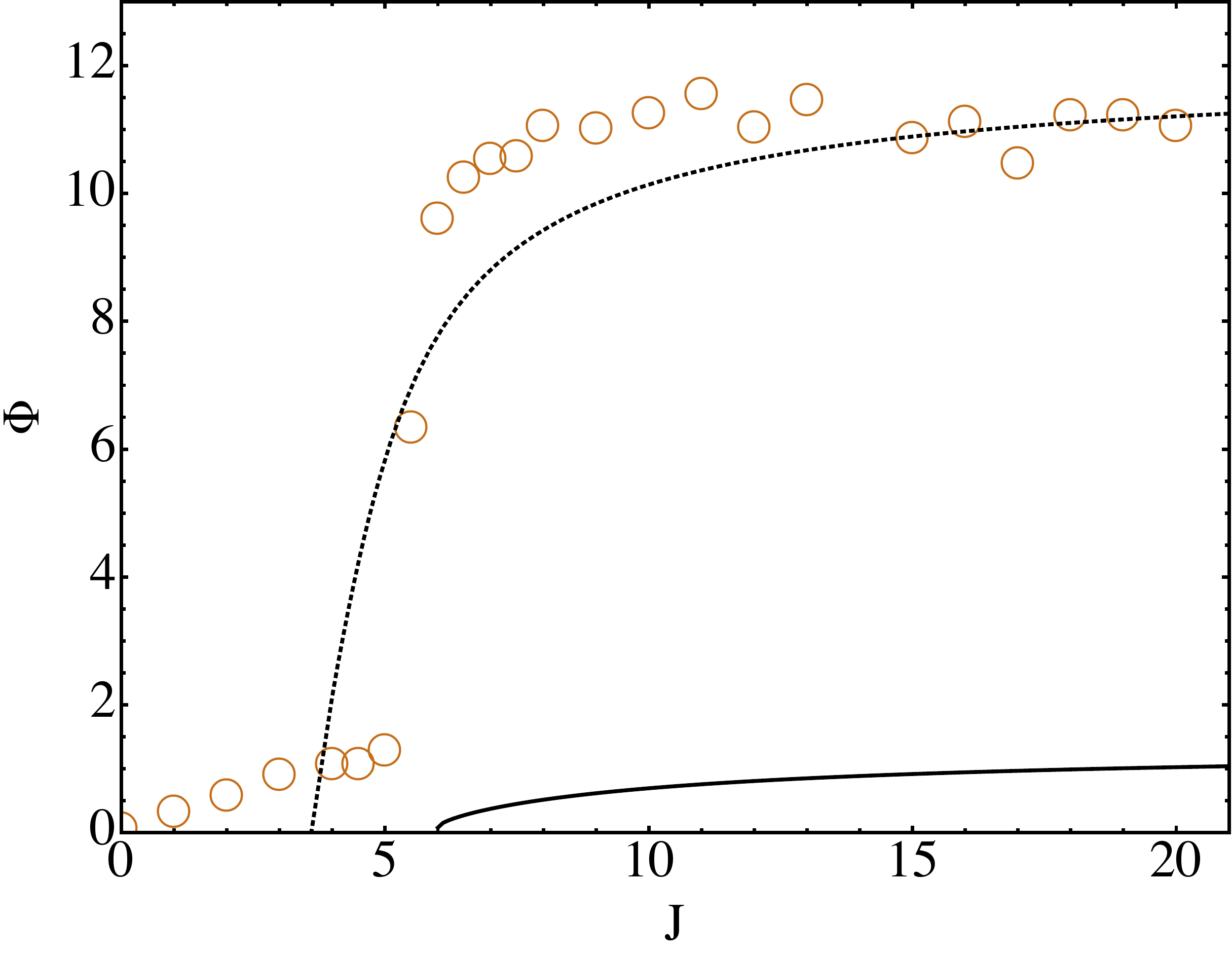}\hspace{0.35cm}
\includegraphics[width=85mm]{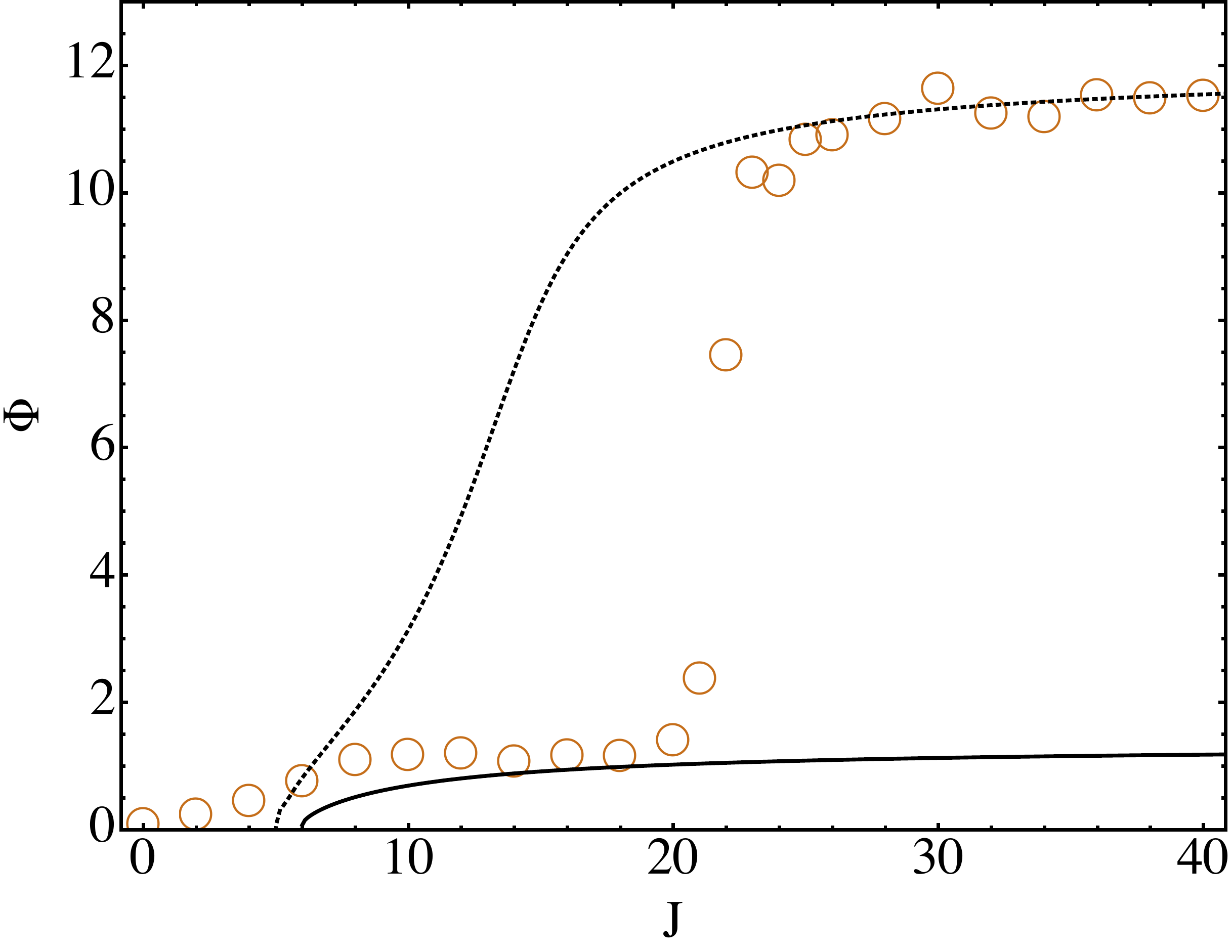}
\includegraphics[width=84mm]{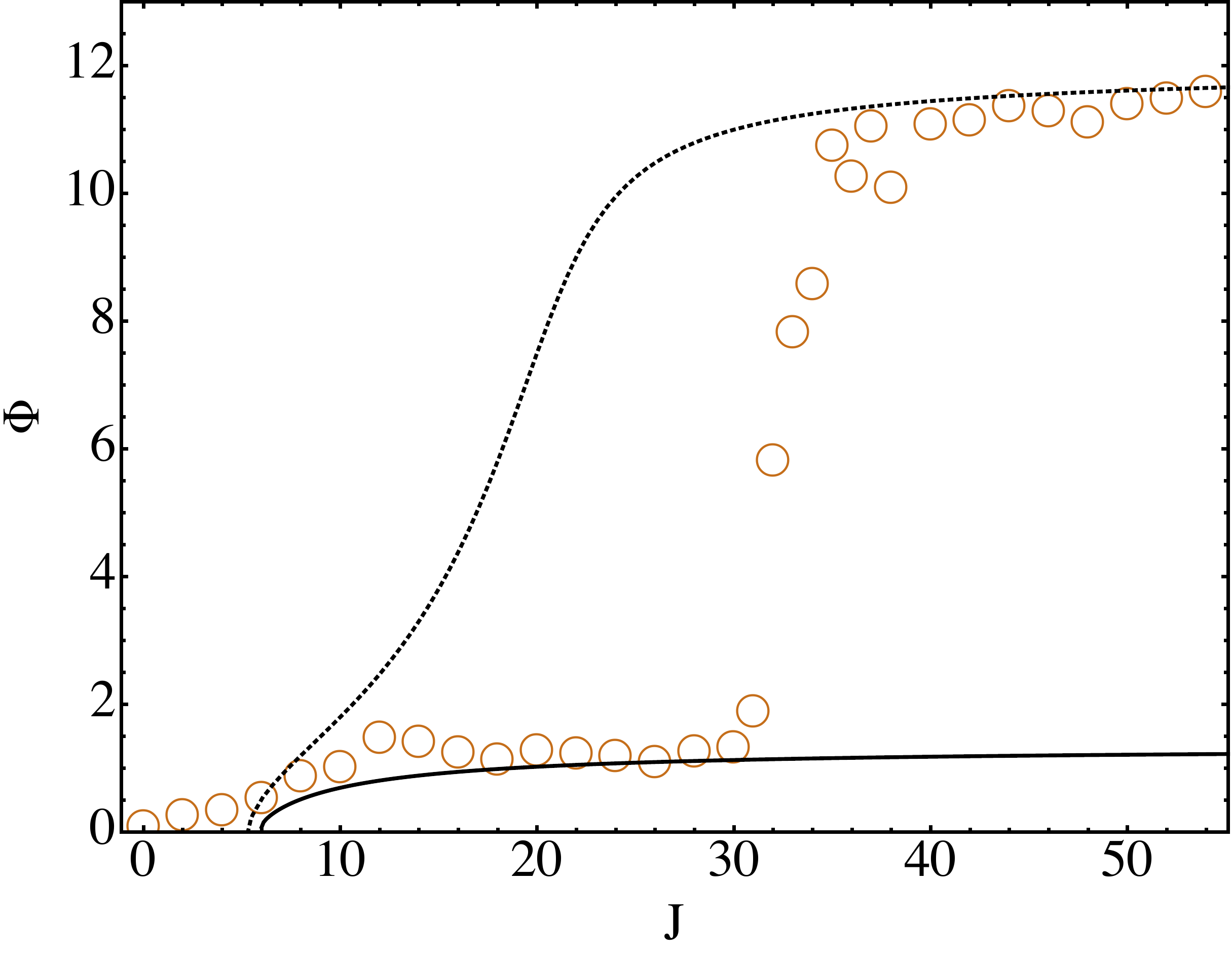}\hspace{0.35cm}
\includegraphics[width=85mm]{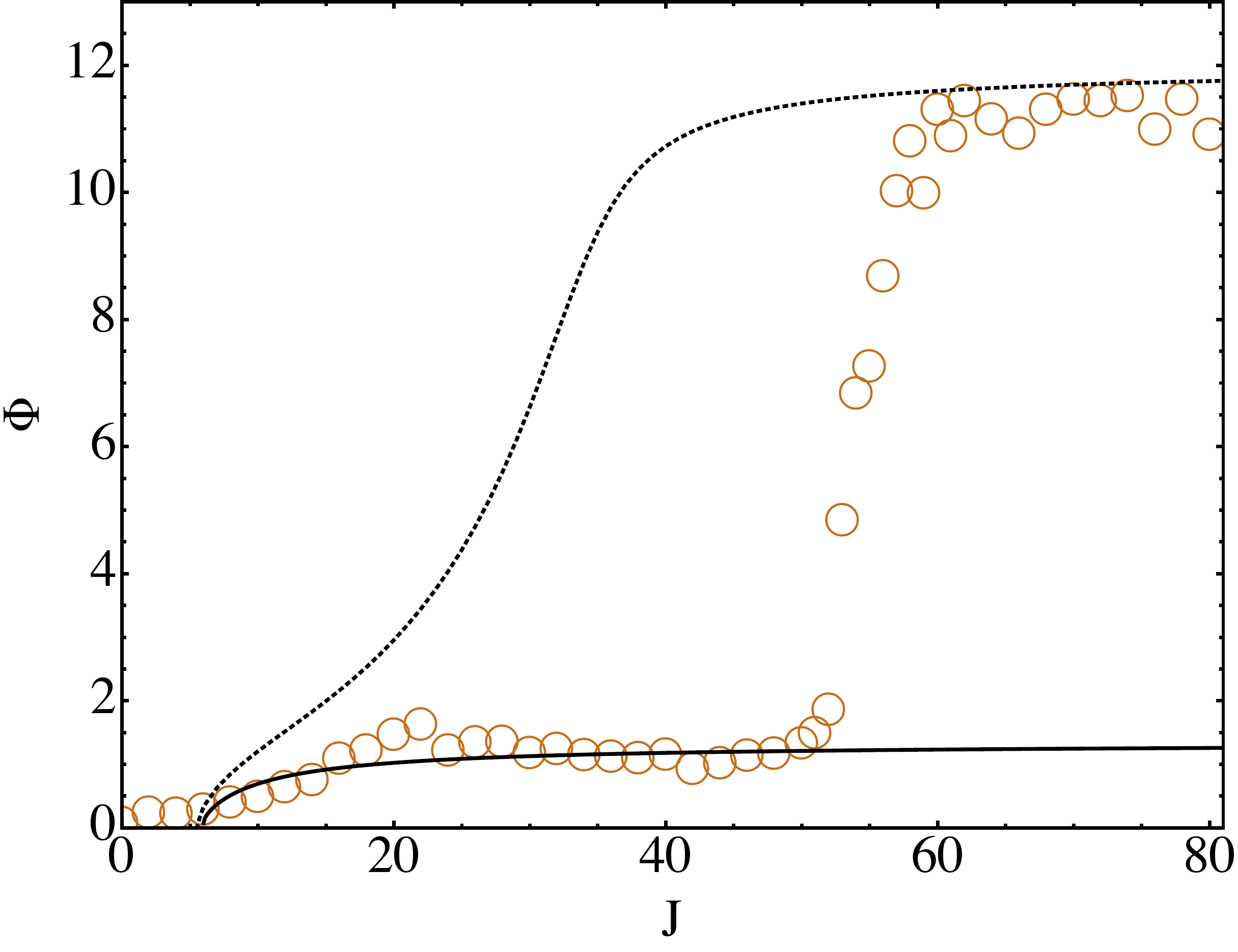}
\begin{picture}(0,0) 
\put(-470,360){\textbf{(a)}}
\put(-210,360){\textbf{(b)}}
\put(-470,170){\textbf{(c)}}
\put(-210,170){\textbf{(d)}}
\end{picture}
\caption{{Observable $\Phi$ as a function of $J$ at fixed $u_1=\{10,40,60,100\}$ for (a), (b), (c), (d), respectively. Dashed and solid black lines correspond to $\Phi$ calculated from the saddle point solutions, assuming either the trivial or topological phase respectively. Yellow points are MC data taken on system size $L=12$.} \label{f:Phi}}
\end{figure}
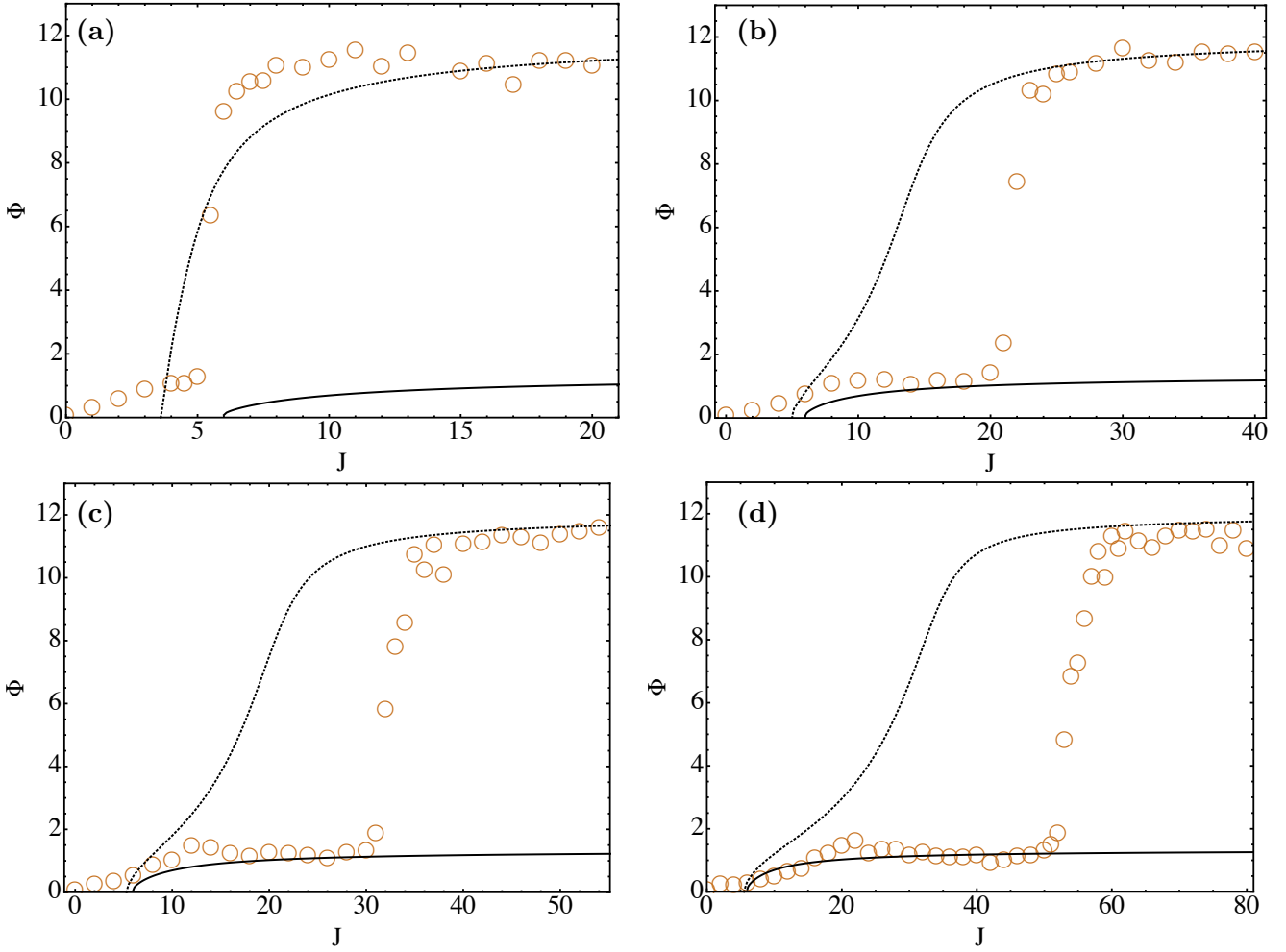

In Figure \ref{f:Phi} we plot the scalar $\Phi$ and its evolution with $J$ for various $u_1$. The data is for collected from simulations with $L=12$. $\Phi$ is also calculated from the saddle point equations by taking the Higgs field configuration of the trivial or topological phases and computing at arbitrary $(u_1,J)$. Comparing the MC data with analytic results, we see quantitative agreement for $\Phi$ deep within each of the topological and trivial Higgs phases, i.e. away from the transition. As already observed from the eigenavlue analysis, the phase numerical $N_h=4$ and analytical $N_h\to\infty$ phase boundaries do not match. Hence we cannot compare the two approaches in this vicinity. Reassuringly, the phase boundaries, as identified via  $\braket{\omega_i}$ and $\Phi$, are indeed consistent. 

We conclude this section by stating that both observables give the same estimate for the topological-to-trivial phase boundary. And that the combined results of (i) and (ii) paint a convincing picture of the underlying phases and transitions.

\section{Discussion}\label{S:discussion}

We study phases of $SU(2)$ gauge theory with multiple adjoint Higgs fields in $2+1$ dimensions. Such a gauge theory has been motivated physically as a theory for optimal doping criticality in the cuprate superconductors \cite{SSST19}, whereby the confining phase corresponds to the Fermi liquid, while the Higgs phases (both topological and trivial) are the candidates for pseudogap phase.   

The primary motivation of the present work was to determine whether the phases of interest physically -- the confining (Fermi liquid), trivial and topological Higgs (pseudogaps) -- are stable and survive at strong gauge field coupling. To investigate, we employ two complementary approaches; an analytic saddle point analysis, which relies on a large number of Higgs flavours ($N_h\to\infty$), and a numerical Monte Carlo analysis at the physically relevant $N_h=4$ Higgs flavours. We demonstrate that all three phases are stable and occupy a non-zero volume in the phase diagram. The results lend support to the $SU(2)$ gauge theory with multiple adjoint Higgs fields as a candidate low energy description of the optimally doped cuprates. Moreover, the agreement between the $N_h\to\infty$ saddle point analysis and that of the numerical $N_h=4$, suggests that $O(1/N_h)$ corrections do not destabilize the phase diagram. This finding also serves as a consistency check for future large $N_h$ analytic studies of this model. 

Aside from the original physical motivation, the present work has established that the minimal model (\ref{S0}), which is obtained from just the first order expansion in the strong gauge coupling expansion (\ref{Dlink}), is sufficient to generate a stable $\mathbb{Z}_2$ topologically ordered phase. We expect that such generic minimal models will also be applicable in the context of spin-liquids.

\subsection*{Acknowledgements}
%********************************************************

We thank Ying Jer Kao for valuable discussions.
This research was supported by the National Science Foundation under Grant No. DMR-1664842. 
MS acknowledges support from the German National Academy of Sciences Leopoldina through grant LPDS 2016-12. HS acknowledges support from the Australian-American Fulbright Commission.

%\bibliography{su2}


\newpage

\appendix
\section{Strong-Coupling Expansion: Higgs and gauge plaquettes}
We denote gauge plaquettes
\begin{align}
\square_{(\bar\theta_1,\bar\theta_2,\bar\theta_3,\bar\theta_4)}\equiv \Tr\left(U(\bar\theta_1)U(\bar\theta_2)U(\bar\theta_3)U(\bar\theta_4)\right),
\end{align}
and we find that only even powers integrate to non-zero
\begin{align}
\braket{\square_{(\bar\theta_1,\bar\theta_2,\bar\theta_3,\bar\theta_4)}}_U&=0\\
\braket{\square^2_{(\bar\theta_1,\bar\theta_2,\bar\theta_3,\bar\theta_4)}}_U&=1.
\end{align}

Let's consider Higgs-hopping terms decorating an elementary plaquette, and expand in Gauge plaquettes.
The lowest-order terms that are trivial in adjoint indices are 
\begin{align}
\label{HHGexp}
\braket{{\cal U}_{ab}(\bar\theta_1){\cal U}_{bc}(\bar\theta_2){\cal U}_{cd}(\bar\theta_3){\cal U}_{da}(\bar\theta_4)}_U&=0\\
\braket{{\cal U}_{ab}(\bar\theta_1){\cal U}_{bc}(\bar\theta_2){\cal U}_{cd}(\bar\theta_3){\cal U}_{da}(\bar\theta_4) \ \square_{(\bar\theta_1,\bar\theta_2,\bar\theta_3,\bar\theta_4)}}_U&=0\\
\braket{{\cal U}_{ab}(\bar\theta_1){\cal U}_{bc}(\bar\theta_2){\cal U}_{cd}(\bar\theta_3){\cal U}_{da}(\bar\theta_4) \ \square^2_{(\bar\theta_1,\bar\theta_2,\bar\theta_3,\bar\theta_4)}}_U&=\frac{1}{16}\\
\braket{{\cal U}_{ab}(\bar\theta_1){\cal U}_{ab}(\bar\theta_1) \ \square^2_{(\bar\theta_1,\bar\theta_2,\bar\theta_3,\bar\theta_4)}}_U=\braket{{\cal U}_{ab}(\bar\theta_1){\cal U}_{ab}(\bar\theta_1)}_U \braket{\square^2_{(\bar\theta_1,\bar\theta_2,\bar\theta_3,\bar\theta_4)}}_U&=\frac{3}{4}.
\end{align}
The final expectation value should will not contribute to the expansion since it will be cancelled by the disconnected vacuum.  The contribution to the action is then 
\begin{align}
\label{Plink}
\braket{\hat{P}}_U&=\frac{1}{16}\frac{\kappa^4}{4!}\frac{2}{2}\frac{\beta^2}{2^2}\sum_{\braket{ijkl}\in\square}H_{ah}(i)H_{al}(i)H_{bl}(j)H_{bm}(j)H_{cm}(k)H_{cn}(k)H_{dn}(l)H_{dh}(l)\ ,
\end{align}
where $\hat{P}$ is shown diagrammatically in Figure \ref{f:diagrams}.
Factors: $1/16$ from average; $\kappa^4/4!$ from fourth order expansion of Higgs-Gauge links; $2$ from two directions around a single plaquette; $1/2$ from second order expansion of the Gauge-plaquette; $(\beta^2/2)^2$ due to definition of Gauge-plaquette coupling constant. We caution the reader that the prefactor obtained in (\ref{Plink}) assumes a particular (clockwise) orientation of $\braket{ijkl}\in\square$ on the plaquettes.

\section{Saddle Point Solutions}\label{A:saddle}
\subsection{Disordered}
\subsubsection{Class II}
We also search for $A_0\neq0$ solutions. We make use of the rearrangement 
\begin{align}
(\bar\lambda+B_0)\frac{1}{3} - \frac{6}{J}A_0^2&=\int_{-\pi}^\pi\frac{d^3k}{(2\pi)^3} (\bar\lambda+B_0 - 6 A_0 \cos k_x)G(k)=1\\
\Rightarrow \ \ \bar\lambda+B_0 &=3\left(1+\frac{6}{J}A_0^2\right)
\end{align}
Substituting into (\ref{soa}) and (\ref{sob}),
\begin{align}
1&=\int_{-\pi}^\pi\frac{d^3k}{(2\pi)^3} \frac{3}{3\left(1+\frac{6}{J}A_0^2\right)-6A_{0}+A_0(6-2\sum_\mu\cos k_\mu)}\\
A_0&=J\int_{-\pi}^\pi\frac{d^3k}{(2\pi)^3} \frac{\cos k_x}{3\left(1+\frac{6}{J}A_0^2\right)-6A_{0}+A_0(6-2\sum_\mu\cos k_\mu)}.
\end{align}
We now manipulate 
\begin{align}
A_0&=\int_{-\pi}^\pi\frac{d^3k}{(2\pi)^3} \frac{3}{\beta+(6-2\sum_\mu\cos k_\mu)}\\
A^2_0&=J\int_{-\pi}^\pi\frac{d^3k}{(2\pi)^3} \frac{\cos k_x}{\beta+(6-2\sum_\mu\cos k_\mu)}\\
\beta&\equiv \frac{1}{A_0}\left[3\left(1+\frac{6}{J}A_0^2\right)-6A_{0}\right]
\end{align}
And solve numerically for $\beta$, using 
\begin{align}
\left(\int_{-\pi}^\pi\frac{d^3k}{(2\pi)^3} \frac{3}{\beta+(6-2\sum_\mu\cos k_\mu)}\right)^2=J\int_{-\pi}^\pi\frac{d^3k}{(2\pi)^3} \frac{\cos k_x}{\beta+(6-2\sum_\mu\cos k_\mu)}.
\end{align}
Finally, having obtained $\beta$ as a function of $J$, we invert the definition to find $A_0(\beta,J)$, i.e.
\begin{align}
A_0^\pm(\beta,J)&=\frac{1}{36} \left((\beta +6) J\pm\sqrt{(\beta +6)^2 J^2 - 216 J}\right).
\end{align}
We find that the $A_0^+$ solution is inconsistent with the saddle point equations, and so we only keep $A_0^-$. For this solution, $J\in(6.67,9)$, $\beta\in(0,\infty)$ and therefore $0\leq A_0^-<1$. This solution is also independent of $u_1$. 

Inspecting the free energy for {\it class I} and $II$,
\begin{align}
f_I&=\frac{u_1}{6}-\frac{3}{2}(1-\ln3)\\
f_{II}&=\frac{u_1}{6} - \frac{9A_0^2}{2J}-\frac{3}{2}\left(1-\int_{-\pi}^\pi\frac{d^3k}{(2\pi)^3}\ln\left(3\left(1+\frac{6}{J}A_0^2\right)-6A_{0}+A_0(6-2\sum_\mu\cos k_\mu)\right)\right),
\end{align}
we see (via numerical evaluation) that the difference is positive for all $A_0^-(J)$
\begin{align}
\notag&f_{II}-f_I=\\
\notag & - \frac{9A_0^-(J)^2}{2J} +\frac{3}{2}\int_{-\pi}^\pi\frac{d^3k}{(2\pi)^3}\ln\left(3\left(1+\frac{6}{J}A_0^-(J)^2\right)-6A_0^-(J)+A_0^-(J)(6-2\sum_\mu\cos k_\mu)\right)- \frac{3}{2}\ln3 > 0.
\end{align}
This holds for all $u_1$ -- hence only {\it class I} is found in the phase diagram spanned by $(u_1, J)$.

\subsection{Topological}
 \subsubsection{Class II} 
 The second class of solution has $H_{01}\equiv H_1$ and $H_{02}=H_{03}\equiv H_2$, which gives
 $A_{01}=A_1$ and $A_{02}=A_{03}= A_2$ and similarly $B_{01}=B_1$ and $B_{02}=B_{03}= B_2$. We reduce the saddle point equations to expressions in $\bar\lambda$ only,
\begin{align}
A_{\pm}&=\frac{1}{4\left(3-\frac{u_1}{J}\right)}\left(\bar\lambda \pm \sqrt{\bar\lambda^2+\frac{8}{3u_1}\left(3-\frac{u_1}{J}\right)}\right)\\
B_{\pm}&=6A_{\pm} - \bar\lambda\\
H_{\pm}^2&=\frac{A_{\pm}}{J}-\frac{\gamma_2}{A_\pm}.
\end{align}
Now there are two possibilities: $A_1=A_\pm$, $A_2=A_\mp$. For each case, one finds $\bar\lambda$ analytically by solving 
\begin{align}
\sum_a B_a&=2u_1=6(A_\pm + 2A_\mp) - 3\bar\lambda.
\end{align}
However, we find that one of $A_+$ or $A_-$ is negative for any $J$, and hence the Greens function is negative (since in this phase $m^2=0$) and therefore the logarithm in free energy yields a complex value. We can safely disregard this solution.

 \subsubsection{Class III} 
Another topological solution has $H_{01}=H_{02}\equiv H$, and $H_{03}=0$. (One can also consider nonzero such that $H_{01}\neq H_{02}$, but these don't provide real solutions.) The saddle point equations can be recast in terms of $\bar\lambda$
\begin{align}
H^2&=\frac{A}{J}-\frac{\gamma_2}{A}\\
B&=6A-\bar\lambda\\
B_3&=2u_1-2B\\
A&=\frac{\sqrt{3} \sqrt{3 J^2 \lambda ^2+24 J^2 u_1-8 J u_1^2}+3 J \lambda }{2 (18 J-6
   u_1)}\\
A_3&=0.
\end{align}
Finally we need to solve the equation in a single variable $\bar\lambda$ 
\begin{align}
1=H^2+\frac{2\gamma_1}{A} + \frac{1}{\bar\lambda+B_3}.
\end{align}
This has multiple roots; we keep only the consistent root. %the resulting free energy is plotted in Figure \ref{f:FE}.

\subsection{Disordered}
\subsubsection{Class II}
There exist another set of solutions where $B_{02}\neq B_{03}$. They are arrived at by setting $A_{02}=A_{03}=0$ and the following manipulations
\begin{align}
\label{B1d}
B_1&=-\bar\lambda + 6A_1\\
\label{B2d}
B_{2,3}&=\frac{1}{2}(\bar\lambda\pm \sqrt{\bar\lambda^2+8u_1}) \\
\label{Hd}
H^2&=\frac{A_1}{J}-\frac{\gamma_2}{A_1} \\
\label{intstep}
1&=H^2+\frac{\gamma_1}{A_1}+\frac{\bar\lambda}{2u_1} \\
\label{lambdad}
\bar\lambda&=2u_1\left(1-\frac{A_1}{J}-\frac{1}{6 A_1}\right).
\end{align}
Finally, solving
\begin{align}
1&=H^2+\frac{\gamma_1}{A_1}+\frac{1}{\bar\lambda+B_2}+\frac{1}{\bar\lambda+B_3}
\end{align}
for $A_1$ provides four roots (as before). We do not present the results, but we find that these do not correspond to a lower free energy than the previous solution where $B_{02}= B_{03}$. 

\section{MC Details}\label{MC details}
\subsection{Re-writing the action}
We have the effective action (re-produced here for convenience)
\begin{align}
\label{S0a}
S_0&=-\frac{J}{2N_h}\sum_{\braket{ij}} H_{al}(i) H_{am}(i) H_{bl}(j) H_{bm}(j) + \frac{u_1}{2 N_h}\sum_i H_{al}(i) H_{am}(i) H_{bl}(i) H_{bm}(i)
\end{align}
with fixed length constraint on each site  (\ref{constraint}).
For implementation of a Wolff cluster type update, we find that it is essential to rewrite the Higgs fields as a {\it flavour vector}
\begin{align}
\label{FlavourVector}
\underbar{H}^a&=\begin{pmatrix} H_1^a \\ H_2^a \\ H_3^a \\ H_4^a \end{pmatrix}
\end{align}
such that the action is 
\begin{align}
S_0&=-\frac{J}{2N_h}\sum_{\braket{ij}} \left[\underbar{H}^a(i)\cdot \underbar{H}^b(j)\right]^2 + \frac{u_1}{2 N_h}\sum_i \left[\underbar{H}^a(i)\cdot \underbar{H}^b(i)\right]^2.
\end{align}

\subsection{Ising Projection}
To implement Wolf cluster updates we must generate an effective Ising model. This is achieved through projecting the Higgs flavour vector (\ref{FlavourVector}) onto a randomly oriented unit four vector $\underbar r$
\begin{align}
\label{FlavourProject}
\underbar{H}^a(i)&=\left| \underbar{H}^a(i)\cdot \underbar r\right|\underbar r \epsilon_i^a \sigma_i + \underbar{H}^a(i) - \left(\underbar{H}^a(i)\cdot\underbar r\right)\underbar r =  \bm{\alpha}_i^a \sigma_i 
+ \bm{\beta}_i^a
\end{align}
where 
\begin{align}
\epsilon^a_i&= \frac{\text{sign}\left(\underbar{H}^a(i)\cdot \underbar r\right)}{\text{sign}\left(\underbar{H}^1(i)\cdot \underbar r\right)} \ , \ \ \ \ \sigma_i=\text{sign}\left(\underbar{H}^1(i)\cdot \underbar r\right).
\end{align}
$\sigma_i$ will play the role of the Ising variable, while $\epsilon^a_i$ absorbs the the different signs of the projections for the different gauge components `$a$'. Meanwhile, the new vectors appearing in Eq. (\ref{FlavourProject}) satisfy the following conditions (by design)
\begin{align}
\bm{\alpha}_i^a\cdot \bm{\beta}_j^b=0\ , \ \ \ \ \ \forall a,b,i,j\ .
\end{align}

\subsection{Ising Model}
Substituting the Ising projection (\ref{FlavourProject}) into the action (\ref{S0a}) and dropping all terms without the Ising degree of freedom $\sigma_i$ -- since they will be constants w.r.t. the Wolff cluster updates -- we obtain 
\begin{align}
\label{S_Ising}
S_\text{Ising}&=-\sum_{\braket{ij}} J_{ij}\sigma_i  \sigma_j\ , \ \ \ \ \ J_{ij}=\frac{J}{N_h}\left[\sum_{a,b} \left(\bm{\alpha}_i^a\cdot \bm{\alpha}_j^b\right)\left(\bm{\beta}_i^a\cdot \bm{\beta}_j^b\right)\right]. 
\end{align}
For the Wolff cluster updates, the basic procedure is:
\noindent $\bm 1.$ Randomly generate Higgs flavour vectors $\underbar{H}^a(i)$ for all $a=1,2,3$ and at each site $(i)$.
\noindent $\bm 2.$ Randomly generate $\underbar r$ (which is uniform across the the lattice).
\noindent $\bm 3.$ Calculate the corresponding $J_{ij}$ and the initial values of $\sigma_i$.
\noindent $\bm 4.$ Perform standard Wolff updates on the $\sigma_i$ variables; i.e. cluster growth with probability $P(i,j)=1-e^{-2J_{ij}}$. Perform $N_{MC}$ such growth steps.
\noindent $\bm 5.$ Recalculate $\underbar{H}^a(i)$ and repeat steps $\bm 2$--$\bm 5$. To maintain ergodicity, we also employ standard local metropolis updates.

\end{document}